\documentclass[twocolumn,aps,preprintnumbers,amsmath,amssymb]{revtex4}

\usepackage{xcolor}
\usepackage{color}
\usepackage[]{graphicx}
\usepackage{latexsym}
\usepackage{natbib}
\usepackage{amsmath}
\usepackage[caption=false]{subfig}
\usepackage{multirow}
\usepackage{mathtools}
\usepackage{textcomp}
\usepackage{rotating}
\usepackage{tikz}
 
\usepackage[final]{feynmp}
\begin{document}

\title{The jamming transition is a k-core percolation transition}

\author{Flaviano Morone$^1$, Kate Burleson-Lesser$^1$,
  H. A. Vinutha$^2$, Srikanth Sastry$^2$, Hern\'an A. Makse$^1$}

\affiliation{$^1$ Levich Institute and Physics Department, City College of
  New York, New York, NY 10031\\
 $^2$ Jawaharlal Nehru Center for Advanced Scientific Research, Jakkur Campus, Bengaluru 560064, India}

\begin{abstract}
  
We explain the structural origin of the jamming transition in jammed
matter as the sudden appearance of k-cores at precise coordination
numbers which are related not to the isostatic point, but to the
sudden emergence of the 3- and 4-cores as given by k-core percolation
theory. At the transition, the k-core variables freeze and the k-core
dominates the appearance of rigidity.  Surprisingly, the 3-D
simulation results can be explained with the result of mean-field
k-core percolation in the Erd\"os-R\'enyi network. That is, the
finite-dimensional transition seems to be explained by the
infinite-dimensional k-core, implying that the structure of the jammed pack
is compatible with a fully random network.


\end{abstract}

\maketitle


The jamming transition occurs when granular materials reach a certain
density such that particle motion is prevented \cite{jamming}. At this
point, the system jams into a disordered packing state that can
sustain a non-zero shear stress. The jamming transition is a
ubiquitous phenomenon that occurs not only with grains but also with
other soft materials like emulsions, colloids, and glasses. Finding
the maximum density at which materials can pack in such a disordered
state is a problem with ramifications in optimization theory as well,
since the jammed state can be thought of as a set of solutions to a
large class of constraint satisfaction problems \cite{baule}. Thus,
any insight into the nature of the jamming transition will have
implications for a large number of problems in disciplines ranging
from physics to computer science and mathematics.

Due to these broad implications, a large number of studies have been
devoted to understanding the underlying nature of the jamming
transition. Early work noted that the transition is driven by the
coordination number, or the average number of contacts of the
particles in the contact network \cite{jamming}.  The transition has
been identified with the isostatic point at which all particles in the
packing begin to satisfy force-balanced equations.  Further
theoretical refinements have been developed, including approaches
inspired by the theory of spin-glasses applied to hard-sphere glasses
\cite{zamponi}, and statistical mechanical ensembles of
equally-weighted jammed configurations \cite{edwards,baule}.


Here we show that there is a simpler topological reason underlying the
jamming transition: the transition is dominated by the sudden
emergence of the giant k-core in the network of contacts. The k-core
is a topological invariant of the contact network defined as the
unique largest subgraph with a minimum degree (i.e., coordination
number) of at least $k$. The concept of the k-core was introduced in
the field of social sciences ~\cite{seidman} to quantify social
network cohesion and it has since been found to have a large number of
applications to network science in general. For instance, it can help
to explain how influential spreaders of information viralize
information in a social network \cite{gallos}; the robustness of
random networks~\cite{dorogotsev}; the structure of the
internet~\cite{carmi, hamelin}; the large-scale structure of the brain
~\cite{sporn}; and the collapse of ecosystems \cite{flaviano}.

Related to the concept of the k-core, k-core percolation is a
well-known mathematical problem \cite{dorogotsev} that studies the
sudden emergence of giant k-cores (a k-connected subgraph) as the
network goes through a series of discontinuous transitions of mixed
nature with first- and second-order features, when one increases the
number of links in the network.  For a random Erd\"os-R\'enyi (ER)
network \cite{er-network}, this problem has been analytically solved
by Wormald and collaborators \cite{wormald}. It was shown that
subsequent k-cores appear at well-defined average degrees (average
coordination numbers in the contact network). The giant 2-core,
corresponding to the giant component studied in percolation, appears
gradually at an average degree $c_2=1$ as shown by the classic result
of Erd\"os and R\'enyi \cite{er}.  However, it was shown that for
$k\ge3$, the subsequent giant k-cores appear suddenly through
first-order transitions where the size of the corresponding k-core
jumps from zero to a finite value, usually quite large compared to the
total size of the network. These transitions occur at sharply defined
values of the average degree. For instance, the 3-core appears
suddenly at $c_3\approx3.35$ where the 3-core jumps from zero to an
occupancy (the number of nodes in the network belonging to the k-core
divided by the total number of nodes) of $p_3 \approx
0.27$. Subsequently, the 4-core appears at $c_4 \approx 5.14$ with
occupancy $p_4 \approx 0.43$, while the 5-core appears at $c_5 \approx
6.81$ with occupancy $p_5 \approx 0.55$. Notice that none of these
transitions coincide with the isostatic transitions at $c=2d$
(frictionless) or $c=d+1$ (frictional) for a jammed system in $d$
dimensions. The predictions of k-core percolation are valid in an ER
network, which is an ensemble of nodes in random networks that is
effectively defined in infinite dimensions and ignores all correlations
between contacts.  That is, the dimensionality of the problem does not
appear in the ER formulation and the resulting network is fully
random; therefore, it is a solution obtained in the mean-field
approximation, as it is called in the physics literature.

Here we employ a quasi-static shear protocol to numerically study the
jamming transition of a 3d packing of frictional spheres when the
coordination number increases as the system jams under shear. We
construct the network of contact points and study the emergence of the
giant k-cores in turn. We find that as the shear strain is increased,
the contact networks develop giant k-cores in succession exactly at
the precise values $c_k$ predicted by k-core percolation theory in an
ER network as obtained in \cite{wormald}.  In particular, the
precursor of the jamming transition occurs at $c_3\approx 3.35$ (and
not at the isostatic point $c=4$) with the appearance of the giant
3-core.

The solution of Wormald \cite{wormald} captures very precisely the
location of the average coordination number at which each k-core
appears in the shear jamming numerical data. This result is surprising
since the ER solution is valid in infinite dimensions for a fully
randomized network where the correlations introduced by the finite
size of the particles in 3d are ignored. The agreement between an
infinite-dimensional result and a finite-dimensional 3d simulation
indicates that correlations introduced by the particles' constraints
are irrelevant. Thus, the jamming transition may be a more simple
constraint satisfaction problem than previously thought. Our results
show the close relation to the k-core problem which was extensively
studied in the mathematical literature and fully solved.

We use the jammed packings already obtained in \cite{sri-arxiv} where
an athermal quasi-static shear protocol has been used to produce a
series of packings at different volume fractions that jam under shear
at different values of the shear strain. The system is monodisperse
and composed of $N=2000$ spheres subjected to athermal quasi-static
shear (AQS) deformation.  Particles interact {\it via} a repulsive
harmonic potential.  We perform athermal quasi-static protocol to
obtain sheared configurations at different densities. To implement
shear, we first do an affine transformation of particle coordinates
in small steps
of $\Delta\Gamma = 5\times10^{-5}$,
followed by energy minimization using a conjugate gradient method and
periodic Lees-Edwards boundary condition 
at a shear strain $\gamma = \frac{\Gamma}{L}$.  The initial
configurations at different densities ($\phi = 0.56-0.627$) for
shearing are produced by starting from an initial equilibrated hard
sphere fluid at $\phi = 0.45$; a fast initial compression is effected
using a Monte Carlo simulation until the desired density is reached for
the initial configurations.

The contact network which we generate using shear deformation of
mono-disperse soft spheres using the AQS protocol, is then used as
input to solve for force and torque balance conditions as described
below, allowing for both normal and tangential forces, which
corresponds to friction. Past work \cite{vinutha_natphys} has
demonstrated that sheared frictionless spheres evolve contact
geometries that can support finite stresses if frictional forces are
also present. We estimate the normal and the tangential force
components independently, {\it i.e.,} in the limit of infinite
friction, using the following procedure. For a given contact network,
we write the force and torque balance conditions in compact form as $M
\mid F \rangle = 0$, where $M$ is a $(\frac{D(D+1)}{2})N \times DC)$
matrix, $C$ is the number of contacts and $\mid F\rangle$ is a vector
of size $DC \times 1$, with $3$ for $D=3$, force components ($f^n,
f^{\theta},f^{\phi}$) for each contact. The matrix $M$ is constructed
from the unit vectors between spheres in contact. Using the matrix
$M$, we construct an energy function $E = \langle F\mid M^{T}M \mid F
\rangle$ which we minimize to obtain force balance solutions [15].  We
minimize the energy function by imposing positivity of normal contact
forces, since we treat systems with repulsive interactions only.  The
force scale is set by the magnitude of the initial guess.

\begin{figure}[h!]
\centering
\includegraphics[width=.5\textwidth]{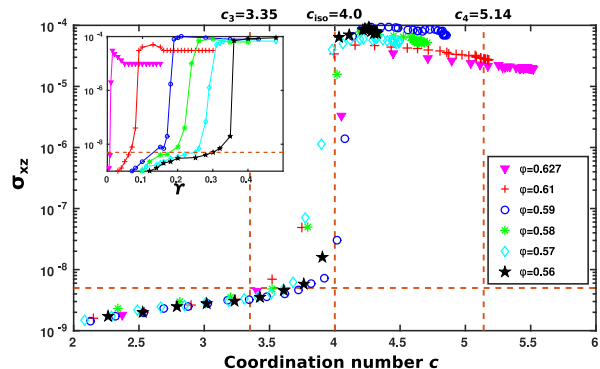}
\caption{Jamming transition. Shear stress versus coordination number
  collapse into a single curve for all volume fractions. Inset shows
  the uncollapsed data when plotted as a function of the shear
  strain. }
 \label{sri}
\end{figure}

The results obtained in \cite{sri-arxiv} can be summarized in
Fig. \ref{sri}.
For all packings examined, a discontinuous jump is seen in the inset
of the figure, in the $xz$-plane shear stress ($\sigma_{xz}$) around
$\sigma_{xz}=5\times10^{-9}$ (indicated by the horizontal dashed line
in both the figure and the inset), at a certain value of the
coordination number $c$ between $c=3$ and $c=4$. The jamming
transition occurs near the vertical line in Fig. \ref{sri} marked
"$c_{\rm iso}=4$," close to the isostatic transition for a
3-dimensional jammed system of frictional particles. This value is
approximately independent of the packing's volume fraction $\phi$;
indeed, the data for all packings collapse roughly to a single curve
for different $\phi$. Although it is close to the expected finding of
a discontinuous jump at $c=d+1$, or $c=4$, this transition appears to
have a precursor, with the onset of the shear stress increase apparent
below the isostatic value of c.
The inset shows $\sigma_{xz}$
as a function of the strain $\gamma$. Again, we see a discontinuous
jump in the shear stress at the jamming transition, occurring at
values of $\gamma$ which are density-dependent and thus unique to each
of the particle configurations examined here.

We analyze these packings to test the idea that k-core percolation is
underlying the sudden transition that is observed in Fig. \ref{sri}.
In fact, the similarities between k-core percolation and the jamming
transition have prompted previous works to propose the conjecture of a
close relation between the two problems \cite{liu}.

\begin{figure}[h!]
\centering
\includegraphics[width=.33\textwidth]{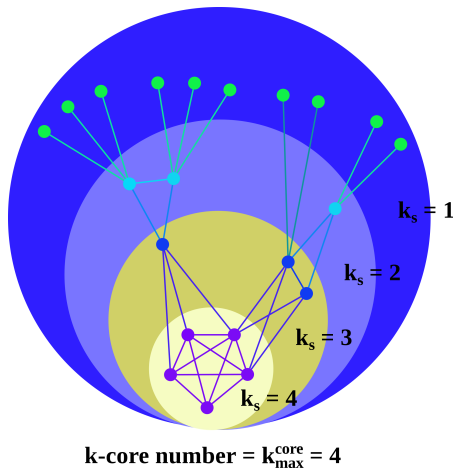}
\caption{Definition of k-core, k-shells and maximum k-core.}
 \label{kcore-definition}
\end{figure}

Figure \ref{kcore-definition} defines the k-core of the network: the
maximal subgraph consisting of nodes having degree at least
$k$~\cite{seidman, dorogotsev}. This subgraph is unique but not
necessarily connected, thus the k-cores might be formed by small
clusters spread around the contact network. An algorithm to extract
the k-core is linear in the system size and consists of iteratively
pruning nodes with degree less than $k$, until the k-core is obtained.
By definition, the k-cores are nested, that is, the k-core contains
the k+1-cores. For instance, the 1-core contains the 2-core, the
2-core contains the 3-core, and so on.  Each k-core is composed of two
structures: the nodes at the periphery called the k-shell and labeled
$k_s$, and the remaining k+1-core.  The periphery of the k-core is
defined as the subgraph induced by nodes and links in the k-core and
not in the k+1-core.  The 1-core corresponds to the full network, and
its connected component is the so-called giant connected component in
percolation.  The 1-shell is a forest, i.e., a collection of
trees. This forest can be removed from the network and the resulting
2-core is also the same as the giant component in percolation in a
statistical sense.  For $k\ge 3$, the k-cores are not related to the
giant component and appear suddenly when we add more links to the
network.  The value $k_{\rm core}^{\rm max}$ of the largest order
k-core, which coincides with the largest value of the k-shell index
$k_s$, is called the k-core number of the network and corresponds to
the innermost core of the network.  It is a topological invariant of
the network, meaning that it does not depend on how the nodes are
labelled or the network portrayed, i.e., it is invariant under
homeomorphisms.

For our analysis, we examine the set of particle configurations with
volume fraction $\phi$; associated with each configuration is a set of
packings with varying coordination numbers $c$ which capture the state
of the configuration before, during, and after the jamming
transition. We begin by constructing an adjacency matrix for each
packing, wherein a value of 1 indicates a contact between two
particles and a value of 0 indicates no contact. A k-shell
decomposition \cite{gallos} is then performed on each matrix,
following the algorithm described above, to determine both the maximum
number of k-shells $k_{\rm core}^{\rm max}$ and the occupancy in each shell from
$k=1$ (the outermost shell) to $k=k_{\rm core}^{\rm max}$ (the innermost
shell, or core).

\begin{figure}[h!]
 \includegraphics[width=.5\textwidth]{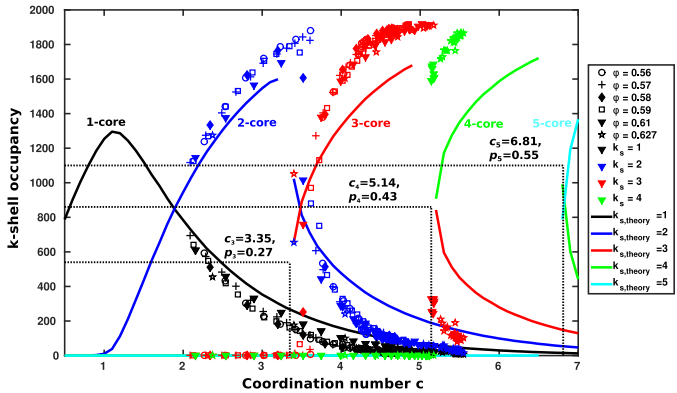} \centering
\caption {The jamming transition is described by the successive
  appearance of giant k-cores at coordination numbers predicted by
  random ER theory \cite{wormald}. }
\label{fig1}
\end{figure}

We find not only that $k_{\rm core}^{\rm max}$ increases with
increasing coordination number $c$ (i.e., those contact networks with
higher coordination numbers have a greater number of k-shells), but
also that there is a rapid transition in the occupancy of $k_{\rm
  core}^{\rm max}$ when a new k-core emerges. As can be seen in
Fig. \ref{fig1}, upon the emergence of a new $k_{\rm core}^{\rm max}$,
the occupancy of the shell that was previously the innermost rapidly
falls to a minimum, while the occupancy of the new innermost
shell\textemdash the new core\textemdash sharply
increases. Furthermore, plotted across all of the networks under
consideration here, these occupancies collapse to a single curve, with
the transition happening at roughly the same point for each packing
network regardless of the packing's volume fraction.

The points at which the new k-cores emerge in the networks of the
packings correspond closely to values theoretically determined via
k-core percolation in random Erd\"os-R\'enyi networks by Wormald
\cite{wormald}. These transition points are indicated in
Fig. \ref{sri} by the vertical lines at $c_3=3.35$, $c_4=5.14$ and
$c_5=6.81$ (for the emergence of the 3-core, 4-core and 5-core,
respectively) and also clearly indicated in Fig. \ref{fig1}.
Furthermore, these data over all packings collapse to a single curve,
regardless of the volume fraction. Notice also that the percolation
transition at $c_2=1$ where the giant component (the 2-core) appears
is irrelevant for the jamming transition, since it appears way before
the larger cores that provide rigidity to the packing.  Indeed,
jamming is described by the appearance of the giant 3-core and not the
giant connected component, which is a tree at the transition point, as
opposed to the 3-core which is a well-connected structure and appears
suddenly rather than continuously like the giant component.  This core
does not appear by nucleation, though. Rather, it appears suddenly,
jumping from zero to a finite fraction of nodes given by $p_3$. This
sudden appearance is due to the fact that jamming requires a global
condition of force balance that is satisfied in all the packing, and
cannot be satisfied by nucleation of specific regions in the
packing. This global feature of jamming might explain the surprising
result of why the physics of jamming is captured by a simple
mean-field infinite-dimensional fully-random non-perturbative k-core
solution even when the packing is three-dimensional.  This result may
not be directly relevant for the glass transition due to the existence
of finite clusters in finite dimensions. In the jamming
zero-temperature description, finite clusters in finite dimensions
violate force balance and then the jamming transition must appear as a
giant core.


When we fully randomized the links in the packings while keeping the
original degree distribution (or coordination number) $c$, we arrive
at a null model\textemdash the Erd\"os-R\'enyi network for this
particular $c$ and number of particles. Even when the correlations are
removed in this manner, it can be seen in Fig.~\ref{fig1} by the solid
lines that the transitions occur at very similar values of $c$ between
the real networks and their fully randomized counterparts.  We use a
set of Erd\"os-R\'enyi networks with the same number of nodes as the
packings ($N=2000$) and average degree equal to coordination number
$c$, for $c=0.5$ to $c=7.0$ in steps of 0.1. Following the procedure
as before, we perform a k-shell decomposition of each network and find
both the occupancy of every shell in the network, and the network's
value of $k_{\rm core}^{\rm max}$. For each value of $c$, 1000 ER
networks are generated, and the values for $k_{\rm core}^{\rm max}$
and shell occupancy are averaged over the generated networks. These
results, shown in Fig.~\ref{fig1} as the solid curves in black; blue;
red; green; and cyan, correspond, respectively, to 1- through
5-cores. At theoretically-predicted coordination numbers $c_k$ and
fractional k-shell occupancies $p_k$, denoted in Fig.~\ref{fig1} by
the dotted lines, the system undergoes transitions wherein a
$k+1$-core (i.e., a new value of $k_{\rm core}^{\rm max}$) emerges
\cite{wormald}. As before, for $k>2$-cores the occupancy of the former
innermost core falls sharply and discontinuously to a minimum while
the occupancy of the new core sharply and discontinuously
increases. It can be seen that the points at which new cores emerge in
the packings match the points at which the new cores emerge in the
generated ER networks, despite the occupancies of the cores being
slightly larger in the packings; this could be the only effect of the
correlations between the particles.  The strong similarity between the
emergence of new cores in both packings and generated ER networks thus
implies that underlying the jamming transition of the packing is the
emergence of a k-core via k-core percolation.


Beyond the jamming transition, the phenomenon of k-core percolation
pertains to other systems whose components (the nodes) require a
minimum number of $k$ connections to other nodes to participate in the
dominant cluster.  Since the k-core sets a constraint on the minimum
number of neighboring nodes, the physics of k-core percolation
describes also the onset of arrested transitions for other systems
with nontrivial constraints, such as spin glasses, glass-forming
liquids, and constraint satisfaction problems (CSP) \cite{mezard}. For
instance, a prototypical model of spin glass systems, known as the $p
\geq 3$-spin glass model, exhibits a critical transition with the same
exponents as k-core percolation, at least at the mean field level.

In the physics of the glass transition, a way to model glassy
dynamics is via kinetically constrained spin lattice models, where
down spins denote regions of low mobility of the liquid, and up spins
denote regions of high mobility.  In addition, a small negative
magnetic field is applied to favor down spins and thus the formation
of clusters of low mobility.  When the temperature of the system is
lowered, more and more clusters of low mobility are formed, eventually
leading to dynamical arrest of the liquid.  The kinetic constraint on
the motion of the spins is such that a spin can flip only if the
number of neighboring up spins is equal to or greater than some
integer $k$, which models the trapping of particles by cages made up
of their neighbors.  Given that up/down spins can be mapped to
present/removed nodes, this kinetic constraint maps to the k-core
condition, and the emergence of a giant cluster of low mobility
regions maps to the k-core percolation.

Another important case is that of constraint satisfaction problems,
where variables must take values which satisfy a number of
constraints.  The random K-XORSAT is an example of such CPS
\cite{mezard}. In this case, the percolation of a 2-core separates the
Easy-SAT and Hard-SAT phase.  In the Easy-SAT phase there is no core,
so that solutions can be found in linear time.
In the Hard-SAT phase there exists a large 2-core, and no algorithm is
known that finds a solution in linear time.  This is due to the
existence of `frozen' variables inside the core (more precisely in the
backbone, which includes the core and all nodes in a corona
surrounding the core), which are fixed in all possible solutions.
Similarly, in the coloring of random graphs, frozen variables appear
if and only if the q-core of the graph is extensive, where $q$ is the
number of colors.

Our results suggest that the onset of jamming in packings can be
understood by the emergence of a 3-core of frozen variables,
analogously to these constraint optimization problems.  Thus, a large
part of the physics of the jamming transition can be explained by this
simple structural picture of the emergence of the 3-core at the
analogous Easy-SAT to Hard-SAT transition.  This is indeed a rigidity
transition when frozen variables appear in the dominating
clusters. After the 3-core has emerged, there is still a hard region
in the coordination number that can be described by following the
phenomenology of more sophisticated spin glass type models such as the
CSP above.

In conclusion, the picture emerging from this study is that the onset
of jamming is not related to the isostatic point nor to regular
percolation but to the sudden emergence of the k-core and that the
structure of the jammed packing is completely random. That is, the
correlations are minimal and the transition is captured well by an ER
network\textemdash an infinitely dimensional network with no
correlations.

\end{document}